\documentclass[aps,pra,twocolumn,floatfix,nopacs,superscriptaddress]{revtex4-1}
\usepackage{amsfonts, amsmath, amssymb}
\usepackage[utf8]{inputenc}
\usepackage[english]{babel}
\usepackage{graphicx}
\usepackage{dcolumn}
\usepackage{braket}
\usepackage{hyperref}
\usepackage{bm}
\usepackage[caption=false]{subfig}
\usepackage{color}
\usepackage{soul}
\usepackage{mathtools}
\usepackage{float}
\newcommand{\be}{\begin{equation}}
\newcommand{\ee}{\end{equation}}
\newcommand{\ba}{\begin{eqnarray}}
\newcommand{\ea}{\end{eqnarray}}

\newcommand{\ada}{\hat{a}^{\dagger}}
\newcommand{\aaa}{\hat{a}}

\newcommand{\ssx}{\hat{\sigma}^x}
\newcommand{\ssy}{\hat{\sigma}^y}
\newcommand{\ssz}{\hat{\sigma}^z}
\newcommand{\ssa}{\hat{\sigma}^\alpha}

\newcommand{\sssx}{\hat{\Sigma}^x}
\newcommand{\sssy}{\hat{\Sigma}^y}
\newcommand{\sssz}{\hat{\Sigma}^z}
\newcommand{\sssa}{\hat{\Sigma}^\alpha}

\def\rme{{\rm {e}}}
\def\rmi{{\rm {i}}}
\renewcommand{\vec}{\mathbf}

  {\left\lbrace\begin{array}{@{}l@{}}}%
  {\end{array}\right.}

\begin{document}

\title{Phase diagram of incoherently driven strongly correlated photonic lattices}

\author{Alberto Biella}
\email{alberto.biella@univ-paris-diderot.fr}
\affiliation{Universit\'{e} Paris Diderot, Sorbonne Paris Cit\'{e}, Laboratoire Mat\'{e}riaux et Ph\'{e}nom\`{e}nes Quantiques, CNRS-UMR7162, 75013 Paris, France}
\affiliation{NEST, Scuola Normale Superiore \& Istituto Nanoscienze-CNR, I-56126 Pisa, Italy}

\author{Florent Storme}
\affiliation{Universit\'{e} Paris Diderot, Sorbonne Paris Cit\'{e}, Laboratoire Mat\'{e}riaux et Ph\'{e}nom\`{e}nes Quantiques, CNRS-UMR7162, 75013 Paris, France}

\author{Jos\'e Lebreuilly}
\affiliation{INO-CNR BEC Center and Dipartimento di Fisica, Universit\`a di Trento, I-38123 Povo, Italy}

\author{\\ Davide Rossini}
\affiliation{Dipartimento di Fisica, Universit\`{a} di Pisa and INFN, Largo Pontecorvo 3, I-56127 Pisa, Italy}
\affiliation{NEST, Scuola Normale Superiore \& Istituto Nanoscienze-CNR, I-56126 Pisa, Italy}

\author{Rosario Fazio}
\affiliation{ICTP, Strada Costiera 11, 34151 Trieste, Italy}
\affiliation{NEST, Scuola Normale Superiore \& Istituto Nanoscienze-CNR, I-56126 Pisa, Italy}

\author{Iacopo Carusotto}
\affiliation{INO-CNR BEC Center and Dipartimento di Fisica, Universit\`a di Trento, I-38123 Povo, Italy}

\author{Cristiano Ciuti}
\affiliation{Universit\'{e} Paris Diderot, Sorbonne Paris Cit\'{e}, Laboratoire Mat\'{e}riaux et Ph\'{e}nom\`{e}nes Quantiques, CNRS-UMR7162, 75013 Paris, France}

\begin{abstract}
We explore theoretically the nonequilibrium photonic phases of an array of coupled cavities in presence of incoherent driving and dissipation. In particular, we consider a Hubbard model system where each site is a Kerr nonlinear resonator coupled to a two-level emitter, which is pumped incoherently.  Within a Gutzwiller mean-field approach, we determine the steady-state phase diagram of such a system.
We find that, at a critical value of the inter-cavity photon hopping rate, a second-order nonequilibrium phase transition associated with the spontaneous breaking of the $U(1)$ symmetry occurs.
The transition from an incompressible Mott-like photon fluid to a coherent delocalized phase is driven by commensurability effects and not by the competition between photon hopping and optical nonlinearity. 
The essence of the mean-field predictions is corroborated by finite-size simulations obtained with matrix product operators and corner-space renormalization methods.
\end{abstract}

\date{\today}
\maketitle

\section{Introduction}
\label{sec:introduction}

Interacting many-particle systems driven away from the thermodynamic equilibrium exhibit a number of interesting features in both the classical~\cite{cross1993,vicsek,zia} and quantum~\cite{petruccione,eisert2015} regimes.
Our understanding of such systems is limited by the fact that nonequilibrium can emerge in a broad variety of forms, making it difficult to develop a unitary picture.
In contrast, in equilibrium situation, very general paradigms allow us to predict the behavior of large ensembles of particles~\cite{fisher1974,fisher1998,pathria,sachdev}.

Nonequilibrium dynamics can be realized both in isolated quantum systems subject to quench dynamics~\cite{polko2011} as well as in open systems under the effect of driving and dissipation.
The investigation of nonequilibrium phenomena in extended open quantum systems is a complex problem: theoretical efforts to develop analytical techniques~\cite{weimer2015,sieberer2016} and numerical methods~\cite{orus2008,mascarenhas2015,cui2015,finazzi2015,montangero2014,ksh2016} have been carried on over the years.
Another strategy involves the experimental realization of a tunable and well controllable {\it quantum simulator}~\cite{feynman,buluta2009} which mimics the behavior of the {\it real} system under consideration.   

The quantum simulation of many-body systems in equilibrium conditions have been implemented in very different contexts ranging from ultracold atoms in optical lattices~\cite{bloch2008,blatt2012} to trapped ions~\cite{monroe2013}.	
The impressive experimental advances of the last decade allowed to extend this idea to the nonequilibrium realm using Rydberg atoms and trapped ions~\cite{muller2012}, exciton-polariton condensates~\cite{carusotto2013}, cold atoms in cavities~\cite{ritsch2013}, and arrays of coupled QED-cavities ~\cite{houck2012,tomadin2010rev}.
Among them, coupled-cavities arrays are particularly appealing from the condensed matter physics perspective since they allow the simulation of archetypal (interacting) lattice models under nonequilibrium conditions~\cite{reviews}.  

\begin{figure}[b!]
\includegraphics[width=1\columnwidth]{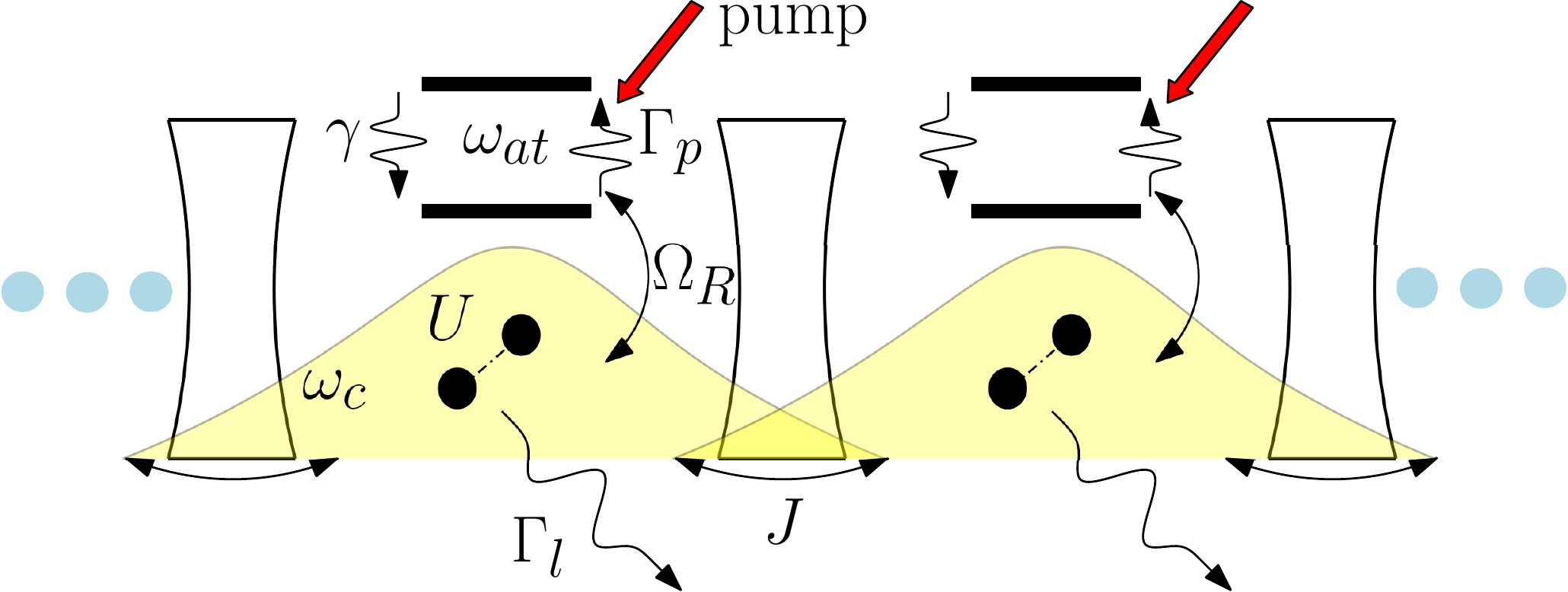}
\caption{(Color online) A sketch of the considered photonic system, consisting of a lattice of coupled nonlinear cavities. Each lattice site is a cavity coupled to a two-level system, which is incoherently pumped at a rate $\Gamma_{p}$. 
$\Omega_R$ is the coherent coupling rate (vacuum Rabi frequency) between the cavity mode and the two-level emitter (with frequency $\omega_c$ and $\omega_{at}$ respectively), while $U$ is
the photon-photon Kerr on-site interaction. The coupling with the environment produces incoherent photon leakage and atomic relaxation at a rate $\Gamma_{l}$ and $\gamma$ respectively. Photons can hop between neighboring sites at a rate $J$.}
  \label{fig:sk_00}
\end{figure}

The emergence of collective behaviors and critical phenomena in these platforms have attracted an increasing interest over the years.
Their phase diagram can be remarkably rich (see for example Refs.~\cite{hartmann2010,lee2011,umucalilar2012,jin2013,lee2013,hoening2014,chan2015,wilson2016,ff2017}): Exotic ordering~\cite{schiro2016} and phases without an equilibrium counterpart may appear~\cite{jin2016}. 
Only very recently, the build up of quantum correlations and the peculiar behavior at criticality started to be investigated~\cite{torre2012,sieberer2013,marino2016,rota2016}.
As it happened for their equilibrium counterpart, the possibility to engineer and manipulate complex many-body states would allow to study fundamental questions and obtain deep insight about the nature of phase transitions and spontaneous out-of-equilibrium ordering. 

An important ingredient determining the dynamics in these systems is the competition between the on-site photon-photon interaction (mediated by the atoms) and the photon-hopping between neighbouring cavities:  
A large local interaction favours the formation of states with a fixed number of particles per site, while a large photon-hopping allows delocalization and enhances the density fluctuations.
The early works in the literature~\cite{hartmann2006,greentree2006,angelakis2007}  explored this physics considering essentially the case of negligible photon losses.
The large number of studies in this regime clearly showed the striking resemblance of the thermodynamic phases~\cite{greentree2006,zhao2008,aichhorn2008,schmidt2009} as well as of the critical properties~\cite{rossini2007,koch2009,hohenadler2011} with the celebrated Bose-Hubbard (BH) model (see also the reviews~\cite{reviews} and the references therein). 

However, the unavoidable presence of photon dissipation and atomic relaxation affects the thermodynamic properties of these quantum simulators.
The dissipative processes are counteracted by an external (coherent or incoherent) driving source which makes the long-time dynamics nontrivial: As a matter of fact, this is determined by the simultaneous interplay between Hamiltonian dynamics, dissipative processes and external driving.
As a result, the scenario considerably enriches and the intrinsic nonequilibrium nature of these platforms emerges in different aspects, ranging from their dynamical response~\cite{tomadin2010} and transport properties~\cite{biella2015,angelakis2015}  to the steady-state behavior~\cite{nissen2012,leboite2013,biondi2016}. 
Furthermore, nonequilibrium effects have been highlighted in single planar cavities, where new complex shapes for the condensate wavefunction have been found and explained~\cite{wouters2008} and novel dispersions for the Goldstone mode predicted~\cite{wouters2007}.
Even if the competition between the photon-hopping and on-site nonlinearity always plays an important role, the analogies with the underlying BH model are in general difficult to be found.

In this work we study the steady-state phases of an incoherently driven photonic lattice (see sketch in Fig.~\ref{fig:sk_00}).
Each site is represented by a lossy nonlinear cavity, where the photon mode is coupled to a two-level system,  which is pumped incoherently~\cite{lebreuilly2015,rivas2014}. 
A recent study~\cite{jose2017} has shown that, via a non-Markovian pump scheme based on a reservoir of two-level systems with engineered spectral features, it is possible to stabilize Mott-like states of photons in spite of the losses. However, a nonequilibrium phase diagram is unknown for this class of systems where a photonic lattice is incoherently-driven. Here we determine the phase diagram within a Gutzwiller mean-field approach including both photon modes and two-level systems, showing the emergence of a second-order phase transition with a $U(1)$ symmetry breaking from a Mott-like incompressible fluid of light to a coherent delocalized phase.
Our Gutzwiller theoretical predictions are consistent with finite-size numerical simulations obtained with two different methods (matrix product operators and corner-space renormalization).  

The paper is organized as follows. In Sec.~\ref{sec:model} we introduce our model, highlighting the main features of the incoherent driving scheme.
In Sec.~\ref{sec:results_mf} we discuss the steady-state phase diagram of the system under the Gutzwiller mean-field approximation. 
In Sec.~\ref{sec:results_ex} we go beyond the Gutzwiller approximation and we show the signatures of our findings in cavity arrays, comparing the mean-field results with finite-size numerical simulations. 
Finally, in Sec.~\ref{sec:conclu} we draw our conclusions and discuss some future perspectives.

\section{The Model}
\label{sec:model}

We consider a driven-dissipative BH model for photons in a $d$-dimensional array of QED-cavities (setting $\hbar=1$)
\begin{equation}
\label{bhmodel}
\hat{H}_{ph} =   \sum_{i} \Bigl( \omega_{c} \ \hat a^\dagger_i  \hat a_i + U \  \hat a^\dagger_i \hat a^\dagger_i \hat a_i\hat a_i \Bigr)  - J 
\sum_{\braket{i,j}}  \hat a_i \hat a^\dagger_{j},
\end{equation}
where $\hat a_i$ ($\hat a^\dagger_i$) are bosonic photon annihilation (creation) operators associated with the $i$-th cavity of the chain with natural frequency $\omega_{c}$, which obey the canonical commutation relations ($\bigl[\hat a_i, \hat a^\dagger_j \bigr] = \delta_{ij} , \bigl[\hat a_i, \hat a_j \bigr] = 0$), $J$ is the hopping rate and $U$ sets the scale of the Kerr nonlinearity. 

Each cavity is coupled to a two-level emitter which is pumped incoherently and provides a driving source for the array.
The atomic evolution and the coupling to the cavities are ruled by 
\begin{eqnarray}
\label{hat_hint}
\hat H_{at} &=&  \omega_{at}\sum_{i}  \hat\sigma_i^{+} \hat\sigma_i^{-}, \cr
&& \cr
\hat H_{I} &=& \Omega_R \sum_{i} \Bigl( \hat a^\dagger_i \hat\sigma_i^{-} +
{\rm H.c.} \Bigr),
\end{eqnarray}
where $\hat\sigma_i^{\pm}=(\ssx_i\pm\rmi\ssy_i)/2$ and $\{ \hat\sigma_i^{\alpha} \ | \ \alpha=x,y,z \}$ are the Pauli matrices acting on the $i$-th site.

The photon leakage from the cavities, atomic relaxation and pumping processes are taken into account by means of a master equation for the density matrix in the Lindblad form 
\begin{equation}
\label{lindblad}
\dot\rho = -{\rm i}\bigl[ \hat H , \rho \bigr] + \mathcal{L}[\rho],
\end{equation}
where $\hat H = \hat H_{ph} + \hat H_{at} + \hat H_{I}$ and 
\be
\label{inco}
\mathcal{L}[\rho] = \sum_{i} \left( \frac{\Gamma_{l}}{2} \mathcal{D}[\aaa_i;\rho] + \frac{\gamma}{2} \mathcal{D}[\hat\sigma_i^{-};\rho] + \frac{\Gamma_{p}}{2} \mathcal{D}[\hat\sigma_i^{+}; \rho]\right), 
\ee
with $\mathcal{D}[\hat{O};\rho] = [ 2\hat{O} \rho\hat{O}^\dagger  -\{   \hat{O}^\dagger  \hat{O},\rho  \} ]$. 
A sketch of the system is provided in Fig.~\ref{fig:sk_00}.
In what follows we will be interested in the nonequilibrium steady-state (NESS) of this model, as determined by computing the long-time limit of Eq.~\eqref{lindblad} $\rho(t\to\infty)=\rho^{SS}$.


\section{Results}
\label{sec:results_mf}
In this section, we present our main results for the steady-state properties of the considered system obtained 
by solving the  the master equation (\ref{lindblad}) via several approaches.
In Sec.~\ref{ssec:single}, we first consider the single-cavity case and discuss the {\it photon number selection mechanism} that can be achieved with the considered incoherent pump scheme for the two-level emitters.
In particular, in the limit of large nonlinearity (hard-core photons), we provide the exact analytical solution for the steady-state density matrix.
In Sec.~\ref{ssec:gutz}, we study the many-cavity case and explore its steady-state phase diagram within a Gutzwiller mean-field approach.
Finally, in Sec.~\ref{sec:results_ex} we compare our findings with exact finite-size simulations using two different techniques.

\begin{figure}[t!]
\centering
\includegraphics[width=1.0\columnwidth]{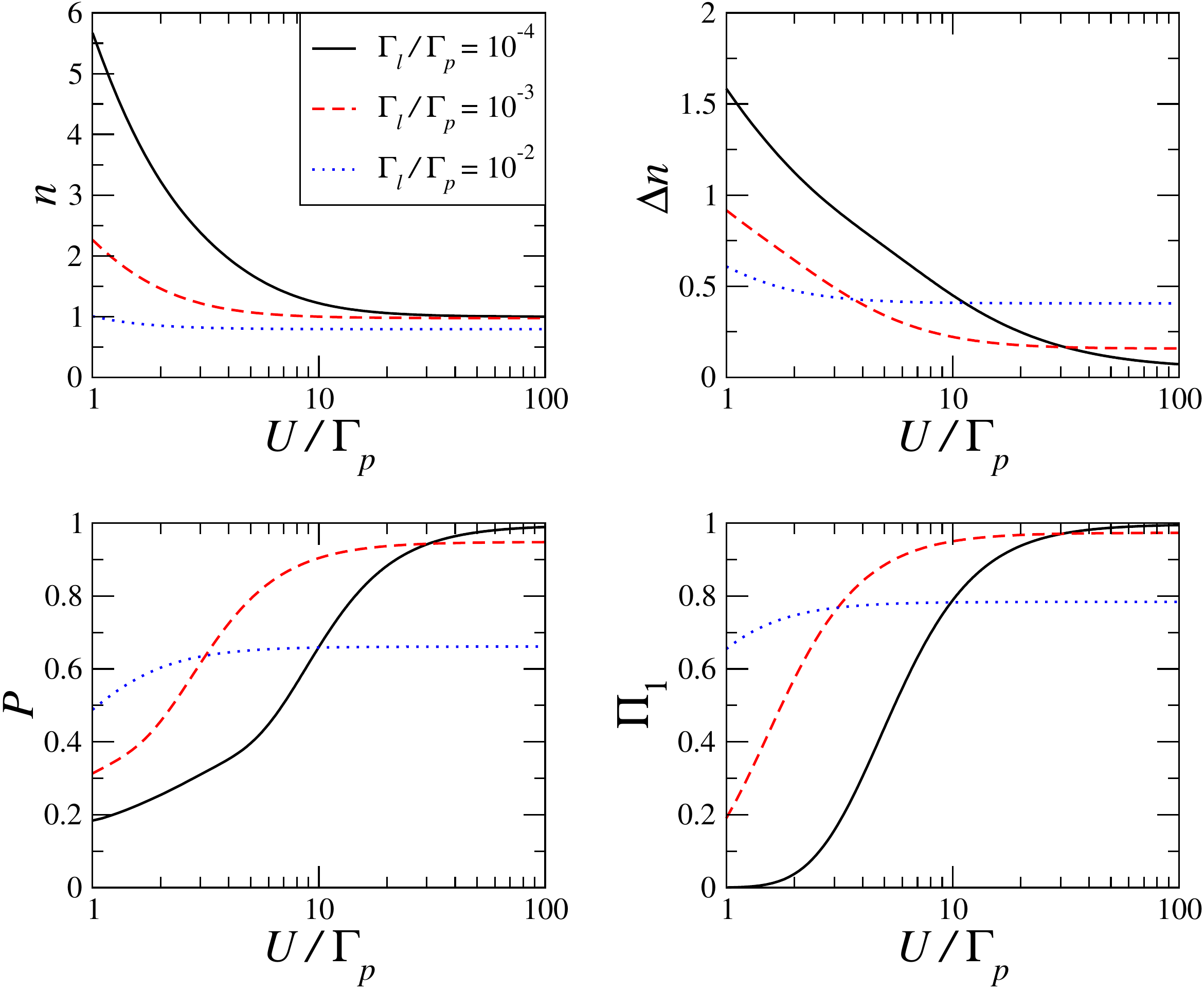}
\caption{(Color online) Steady-state observables for a single-cavity system. The steady-state value of the photon number $n$, photon number fluctuations $\Delta n$, purity $\mathcal{P}$ and one-photon Fock state population $\Pi_1$ are plotted as a function of the nonlinearity $U/\Gamma_{p}$ for different values of the cavity dissipation rate $\Gamma_l/\Gamma_{p}$, as indicated in the legend. 
The other parameters are $\omega_{at}=\omega_{c}$, $\Omega_R/\Gamma_{p}=10^{-1}$ and $\gamma/\Gamma_{p}=10^{-4}$.}
\label{fig:single_cuts}
\end{figure}

\subsection{Photon number selection in a single cavity}
\label{ssec:single}

The eigenvalues of the Hamiltonian in Eq.~\eqref{bhmodel} for $J=0$ (labelled by their photon number) are 
\begin{equation}
\omega_N=N \omega_{c} + N(N-1) U.
\end{equation}
This implies that the $N\to N+1$ transition has a frequency $\omega_{N+1,N}=\omega_{c}+2NU$.
Choosing the emitter transition frequency to be resonant with the $N\to N+1$ transition (i.e.,  $\omega_{at}=\omega_{N+1,N}$), it is possible to obtain a NESS that is a mixed state (with no coherences) dominated by the $N+1$ photon state (and the atom in the excited state). 
While the specific conditions to obtain this for generic integer $N$ require a fine tuning of parameters, as discussed in full detail in~\cite{lebreuilly2015}, the ones for the $N=1$ photon state have the simple form
\begin{equation}
\frac{\Gamma_{em}^0}{\Gamma_{l}}\gg1, \qquad \frac{\Gamma_{em}^0\Gamma_{p}^2}{\Gamma_{l} \ U^2}\ll1,
\end{equation} 
where $\Gamma_{em}^0=4\Omega_R^2/\Gamma_{p}$ .
We checked this (numerically) focusing on the $0\to1$ transition by solving the single-cavity master equation via diagonalization of the corresponding Liouvillian.
In the following we will work in units of $\Gamma_p$.
In Fig.~\ref{fig:single_cuts} we show the steady-state value of the photon density $n=\braket{\ada\aaa}$ (where $\braket{\hat O}={\rm Tr}[\rho^{SS}\hat O]$ and ${\rm Tr}[\rho^{SS}]=1$) and  its variance $\Delta n$ as a function of $U/\Gamma_{p}$ for different values of the cavity dissipation rate $\Gamma_l/\Gamma_{p}$. Moreover, we also show the purity of the density matrix $\mathcal{P} = {\rm Tr}[(\rho^{SS})^2]$ and the population $\Pi_1=\bra{1,\uparrow}\rho^{SS}\ket{1,\uparrow}$, where $\ket{1,\uparrow}$ denotes the state with one photon in the cavity mode and the two-level system into its excited state.
As highlighted in the right bottom panel of Fig.~\ref{fig:single_cuts}, it is possible to prepare the desired Fock state with arbitrary precision for large enough nonlinearity and small photon leakage.

When the nonlinearity is large enough to forbid the double occupation of a lattice site, Eq.~\eqref{lindblad} can be analytically solved for the steady-state (see Appendix~\ref{app:hcsol}).
At resonance ($\omega_{at}=\omega_{c}$) the photon density reads as
\be
\label{hcsol_pop}
n = \frac{4\Gamma_{p}\Omega_R^2}{(\Gamma_{p} +\gamma + \Gamma_{l}) \left( \Gamma_{l} (\Gamma_{p} +\gamma) +4 \Omega_R^2 \right)}.
\ee
Expanding Eq.~\eqref{hcsol_pop} for small effective loss/gain ratio $\eta=\Gamma_l/\Gamma_{em}^0$ we obtain
\be
n = \frac{\Gamma_{p}}{\Gamma_{tot}} -  \frac{\Gamma_{p} + \gamma}{\Gamma_{tot}} \ \eta + \mathcal{O}(\eta^2),
\ee
where $\Gamma_{tot} = \Gamma_{p} + \gamma + \Gamma_{l}$.
The expression above for the photon population tells us that, in the regime of large nonlinearity, it is possible to stabilize in the steady-state a single-photon Fock state for small enough cavity and atomic dissipation rates.
The number of photons fluctuates as
\be
 \Delta  n^2 = \frac{\Gamma_{p} (\gamma+\Gamma_l)}{\Gamma_{tot}^2} -  \frac{(\Gamma_{p} + \gamma)(\Gamma_{p} - \gamma - \Gamma_{l})}{\Gamma_{tot}^2} \ \eta + \mathcal{O}(\eta^2).
\ee
What we want to do next is to study what the effect of a finite coupling between neighbouring cavities ($J\neq0$) is.

\subsection{Gutzwiller phase-diagram for cavity lattices}
\label{ssec:gutz}

In this section, we consider a lattice of cavities, using the same kind of incoherent driving that we have analyzed for the single-cavity case.
Due to the complexity of the problem, we perform a Gutzwiller mean-field approximation~\cite{leboite2013,biondi2016} assuming a factorized ansatz for the global density matrix 
\begin{equation}
\label{ansatz_1site}
\rho_{\text{MF}}=\bigotimes_{i}\rho_{i},
\end{equation}
where $\rho_i$ is the density matrix of the $i$-th site. 
Inserting such ansatz into Eq.~\eqref{lindblad} and assuming the translational invariance ($\rho_i=\rho_j$ , $\forall i,j$) we get an effective master equation of the form
\begin{equation}
\label{lindblad_cmf}
\dot\rho_i = -{\rm i}\bigl[ \hat H_{\text{MF}} , \rho_i \bigr] + \mathcal{L}_i[\rho_i],
\end{equation}
where $\hat H_{\text{MF}}=\hat H_i + \hat H_{\mathcal{B}}$. Here $\hat H_i$ and  $\mathcal{L}_i$ contain all the local terms of the full Hamiltonian $\hat H$ and of the superoperator $\mathcal{L}$ (see Eq.~\eqref{inco}) respectively, acting on the $i$-th site. The term $\hat H_{\mathcal{B}}=- z J (\hat a_{i}^\dagger\braket{\hat a}+\rm{H.c.})$ (where $z$ is the coordination number of the lattice), takes into account the mean-field interactions of the site $i$ with its neighbours.
\begin{figure}[t!]
\centering
\includegraphics[width=0.75\columnwidth]{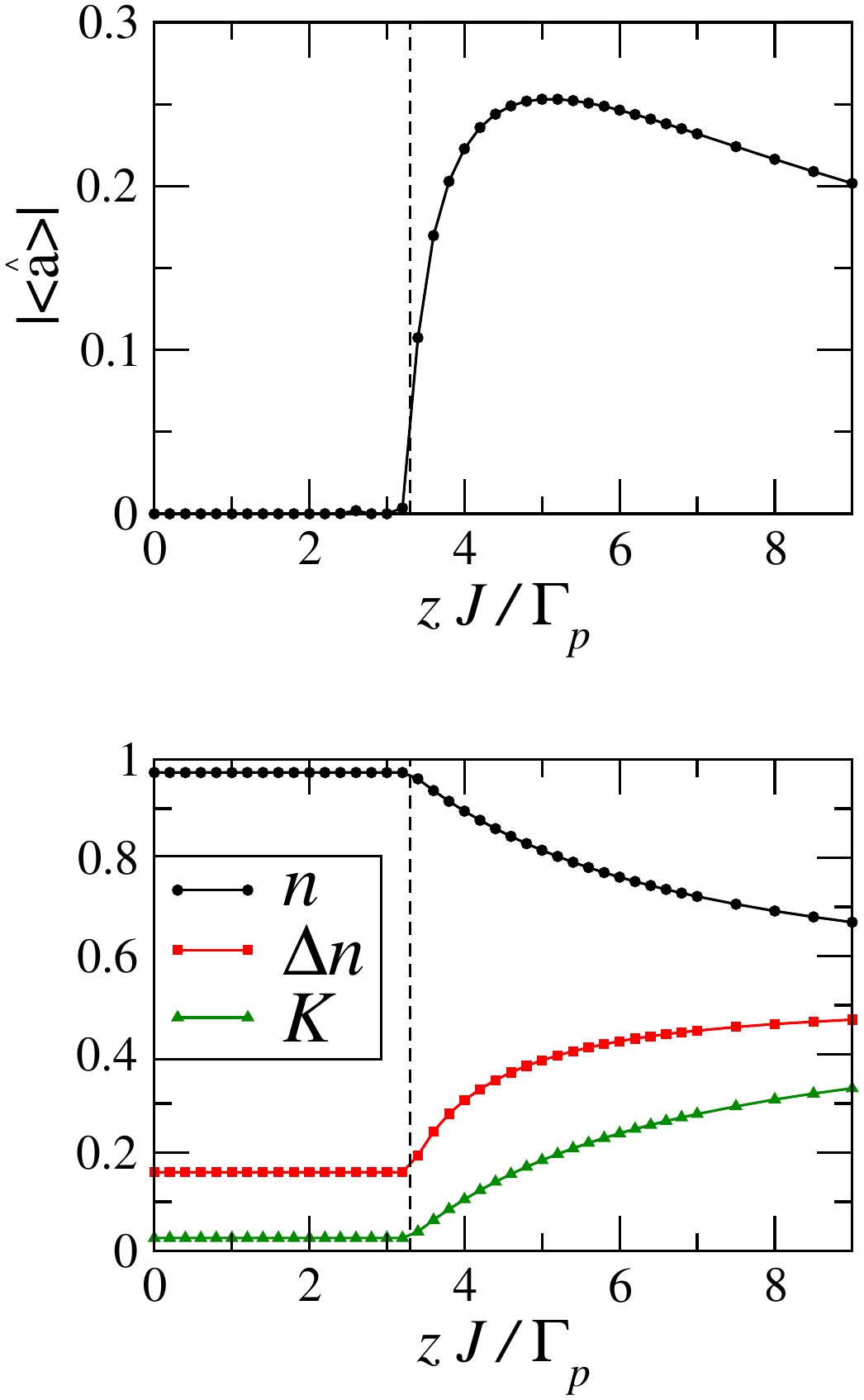}
\caption{(Color online) The order parameter $|\braket{\hat a}|$ (top panel), the number of photons $n$, its variance $\Delta n$ and the compressibility $\mathcal{K}$ (lower panel) of the steady-state of Eq.~\eqref{lindblad_cmf} as a function of $z J/\Gamma_{p}$ in the hard-core limit ($U/J=+\infty$).
Here $\Gamma_{l}/\Gamma_{p}=\gamma/\Gamma_{p}=10^{-3}$, $\Omega_R/\Gamma_{p}=10^{-1}$ and $\omega_{at}=\omega_{c}$.
The dashed vertical line signals the predicted critical value of $J$ (see App.~\ref{app:stab}).}
  \label{fig:MF_1site_sum_bis}
\end{figure}
The steady-state is reached dynamically after integrating Eq.~\eqref{lindblad_cmf} by means of a fourth-order Runge-Kutta method. The mean-field is computed dynamically $\braket{\hat a}(t) = {\rm Tr}[\aaa \ \rho_{i}(t)]$ and used to self-consistently evolve $\rho_{i}(t)$ until the steady-state is reached.
This approach has been proven to be very effective to determine the phase diagram of the
coherently driven BH model~\cite{leboite2013,biondi2016,wilson2016} and similar systems~\cite{jin2013} since the local pumping and decay drastically restrict the range of the correlations.
We therefore expect that the Gutzwiller ansatz~\eqref{ansatz_1site} is well suitable to capture the physics of our system.

In Fig.~\ref{fig:MF_1site_sum_bis} we show the value of the mean-field order parameter $|\braket{\hat a}|$, the photon density $n$, its fluctuations $\Delta n$ and the compressibility $\mathcal{K}=\Delta n^2/n$~\cite{note00} in the steady-state as a function of the inter-cavity hopping rate in the hard-core limit ($U/J=+\infty$). 
The main findings of the Gutzwiller mean-field theory can be summarized as follows. 

For $0\le J<J_c^{\rm HC}$ (where $J_c^{\rm HC}$ denotes the critical hopping rate in the in the hard-core limit), we get a vanishing value of the mean-field order parameter $|\braket{\aaa}|=0$.
As a consequence, each local steady-state density matrix can be approximately written as 
\be
\rho_{i}^{SS}\simeq\ket{1,\uparrow}\bra{1,\uparrow}.
\ee 
By construction, a vanishing value of the MF order parameter forces $\rho_{i}^{SS}$ to be the steady-state solution of the ME~\eqref{lindblad} for a single cavity.
However, the short-range coupling induced by the photon hopping may play an important role. Consequently, in order to characterize the phase with the unbroken symmetry it is necessary to go beyond the mean-field theory. 
This has been done in Sec.~\ref{sec:results_ex} where we show that in a range of $zJ/\Gamma_p$ compatible with the Gutzwiller prediction the number of photons remains very close to the unity with very small fluctuations.
This Mott-like phase, is also characterized by an (almost) vanishing compressibility $\mathcal{K}$,
analogously to what happens in equilibrium situations~\cite{note01}.

At $J  = J_c^{\rm HC}$, a second-order phase transition takes place.
For $J > J_c^{\rm HC}$, the system enters a coherent delocalized phase characterized by the emergence of limit cycles, namely 
\be
\braket{\hat a}=|\braket{\hat a}| \ {\rm e}^{-{\rm i} \omega_L t},
\ee
where $\omega_L$ depends on the system parameters. 
This transition is associated with the spontaneous breaking of the $U(1)$ symmetry possessed by Eq.~\eqref{lindblad}~\cite{note02}.
The steady-state density matrix becomes mixed (not shown) and the photonic population feels the transition: The number of photons starts to be significantly different from $1$ and its fluctuations become relevant.
The transition from a Mott-like to a coherent phase can be measured by monitoring the behavior of the compressibility which becomes finite in the symmetry broken phase.
For large values of $zJ/\Gamma_p$ the photon density remains finite approaching asymptotically $n=1/2$ and the steady-state density matrix becomes maximally mixed. 
We also note that the use of the translationally invariant ansatz~\eqref{ansatz_1site} imposes that the condensate appears in the $\vec{k}=0$ mode~\cite{note03}. This restriction is however fully justified as the pumping conditions chosen for the numerics of Sec.~\ref{sec:results_ex} (see Fig.~\ref{fig:1D_detuned_corner_sum} and Fig.~\ref{fig:1D_detuned_entrderiv}) explicitly favor condensation into the $\vec{k}=0$ mode.

\begin{figure}[t!]
\centering
\includegraphics[width=.75\columnwidth]{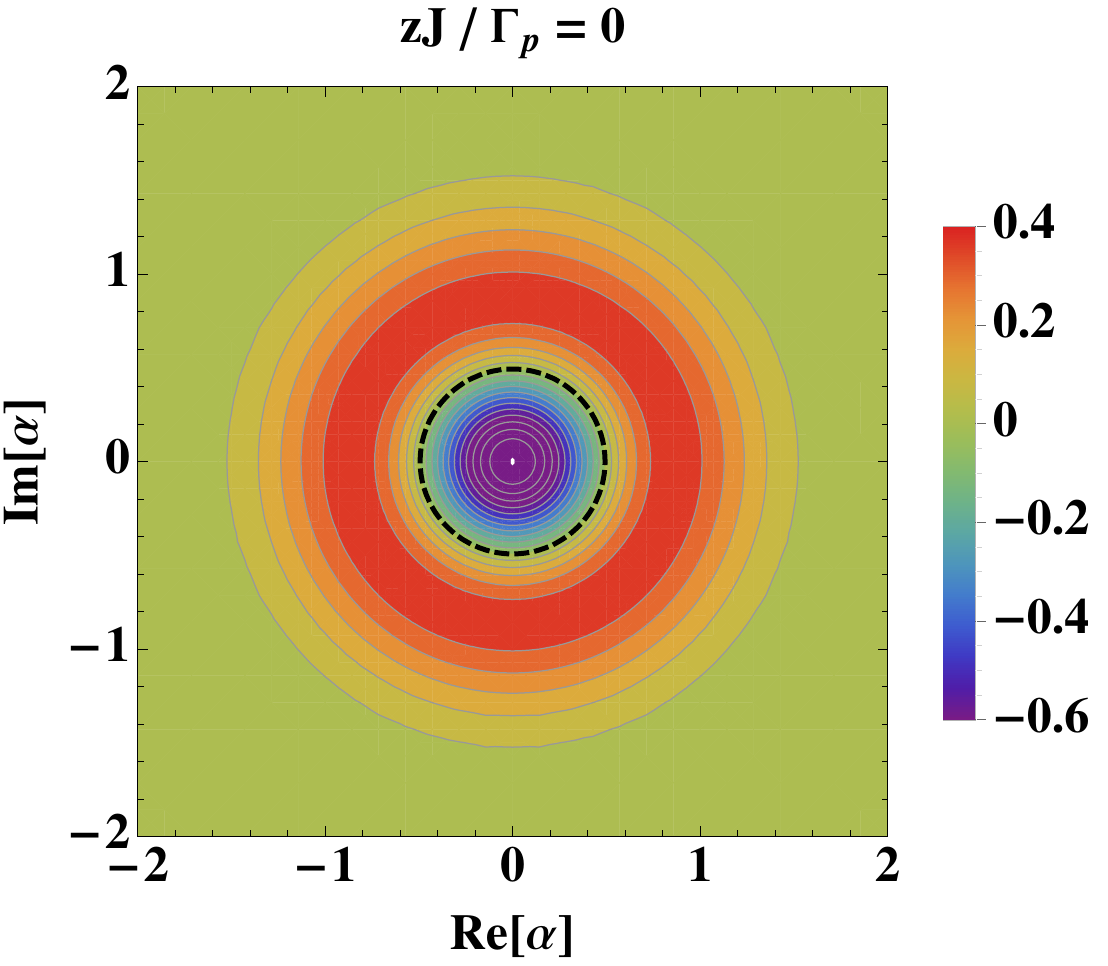} \\
\vspace{+1.0em}
\includegraphics[width=.75\columnwidth]{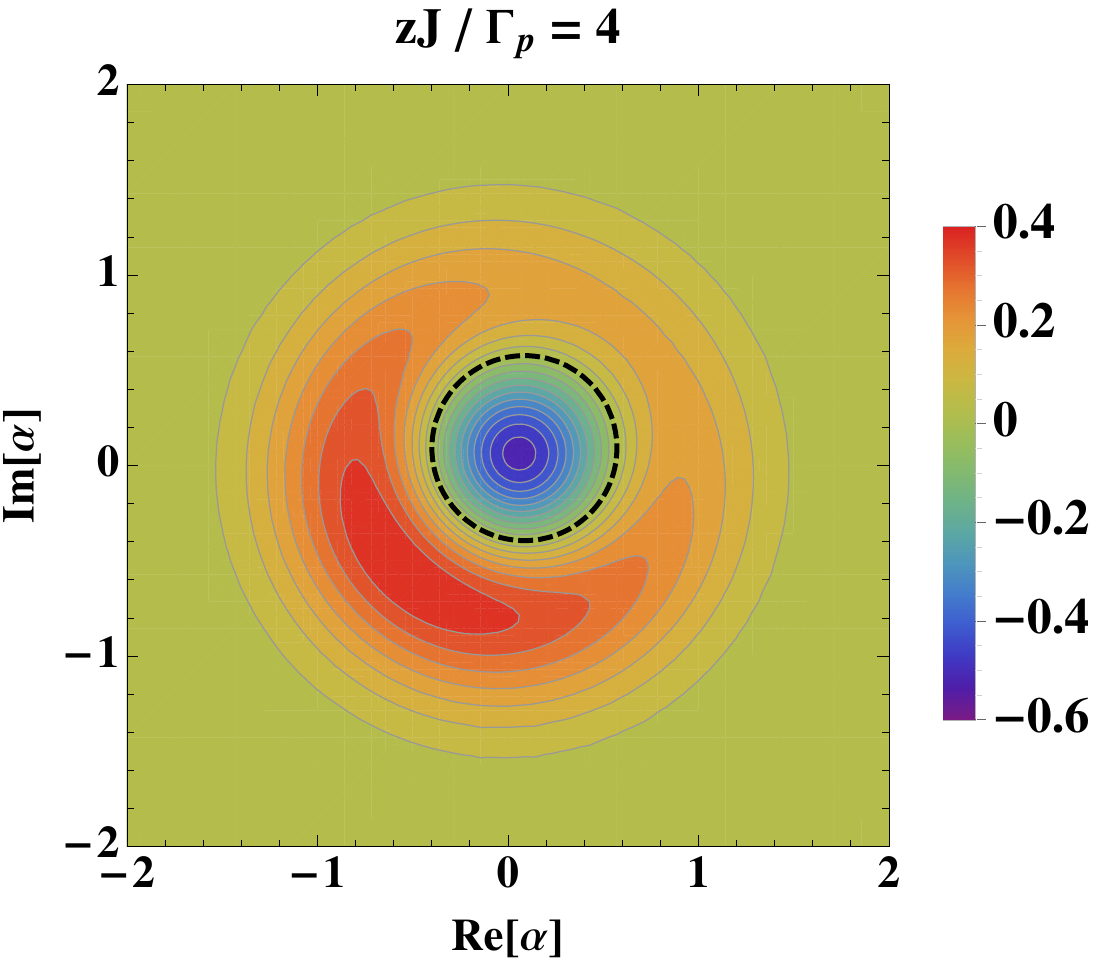}
\caption{(Color online) Contour plot of the steady-state Wigner distribution $W(\alpha)$ in the Mott-like (top panel) and coherent phase (bottom panel).
Each contour denotes a variation of $0.05$ of the value of $W(\alpha)$. 
The black dashed contour encircles the region with $W(\alpha)<0$.
The parameters are set as in Fig.~\ref{fig:MF_1site_sum_bis}.}
\label{fig:MF_wig}
\end{figure}

Further information about the nature of the steady-state can be extracted by looking at the Wigner quasi-probability distribution 
\be
W(\alpha) = \frac2\pi \ {\rm Tr}\left[ \rho^{SS}_{ph} \hat{D}(\alpha) \rme^{\rmi \pi \ada\aaa} \hat{D}^\dagger (\alpha)\right],
\ee
where $ \rho^{SS}_{ph}={\rm Tr}_{at}\left[ \rho^{SS} \right]$ is the photonic reduced density matrix, $\hat{D}(\alpha)=\rme^{\alpha \ada - \alpha^*\aaa}$ is the displacement operator and the prefactor ensures that $\int_{\mathbb{C}}d^2 \alpha  \ W(\alpha)=1$.
As it is shown in Fig.~\ref{fig:MF_wig}, for $0\leq J<J_c^{\rm HC}$ the steady-state is approximately a one-photon Fock state and then the Wigner distribution is negative around $\alpha=0$ (see top panel) indicating strong nonclassicality~\cite{haroche_book}.
For $J>J_c^{\rm HC}$ the $U(1)$ symmetry is spontaneously broken in the steady-state. The Wigner function thus becomes asymmetric and rotates at a frequency $\omega_L$ around $\alpha=0$ (see bottom panel).
Let us note that also in this case $W(\alpha)$ is negative around $\alpha=0$ again indicating the nonclassical nature of the steady-state at low photon number per site.

The phase transition we observe is not related to the competition between photon hopping and nonlinearity since we explored a range of parameters such that the $zJ/U$ ratio remains very small. 
Increasing $J$ we increase the bandwidth ($2zJ$), therefore the photons start to be off-resonant with respect to the incoherent driving source provided by the atoms. 
As a consequence, the photons cannot be efficiently pumped into the array and their number is no longer commensurate to the system size.
The presence of a significant number of empty sites allows photons to move along the lattice and to establish a long-distance coherence.
As we will discuss in Sec.~\ref{sec:results_ex}, the transition can be characterized also in terms of spatial correlation functions: 
it is associated to the appearance of a long-range order in the first-order coherence function. 
Very recently, the transition between a Mott-like regime to a coherent one driven by the $J-U$ competition has been theoretically investigated in~\cite{jose2017} by making use of specifically designed emission spectra based on a non-Markovian reservoir. 

Also for the configuration under exam in the present work, in spite of its nonequilibrium nature it is possible to trace an analogy with the phase diagram of hard-core BH model at zero temperature~\cite{schmid2002}.
The ground-state of this model is a superfluid (in two dimensions) when the band is not completely filled (i.e., $|\mu/J|<z$, where $\mu$ is the chemical potential). 
Here, this commensurability driven transition is extended to the nonequilibrium realm and the emergent coherent phase ($|\braket{a}|\neq0$) has a mixed nature.

\begin{figure}[t!]
\centering
\subfloat[][]
{\includegraphics[width=.44\columnwidth]{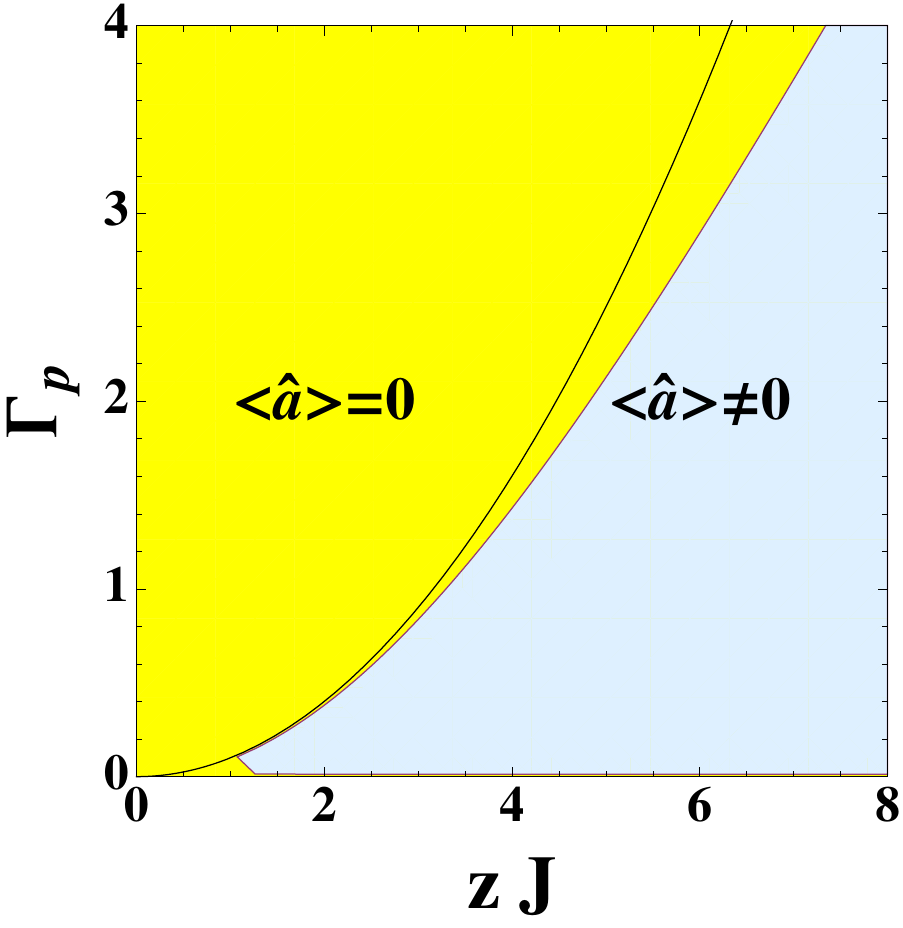}} \quad
\subfloat[][]
{\includegraphics[width=.47\columnwidth]{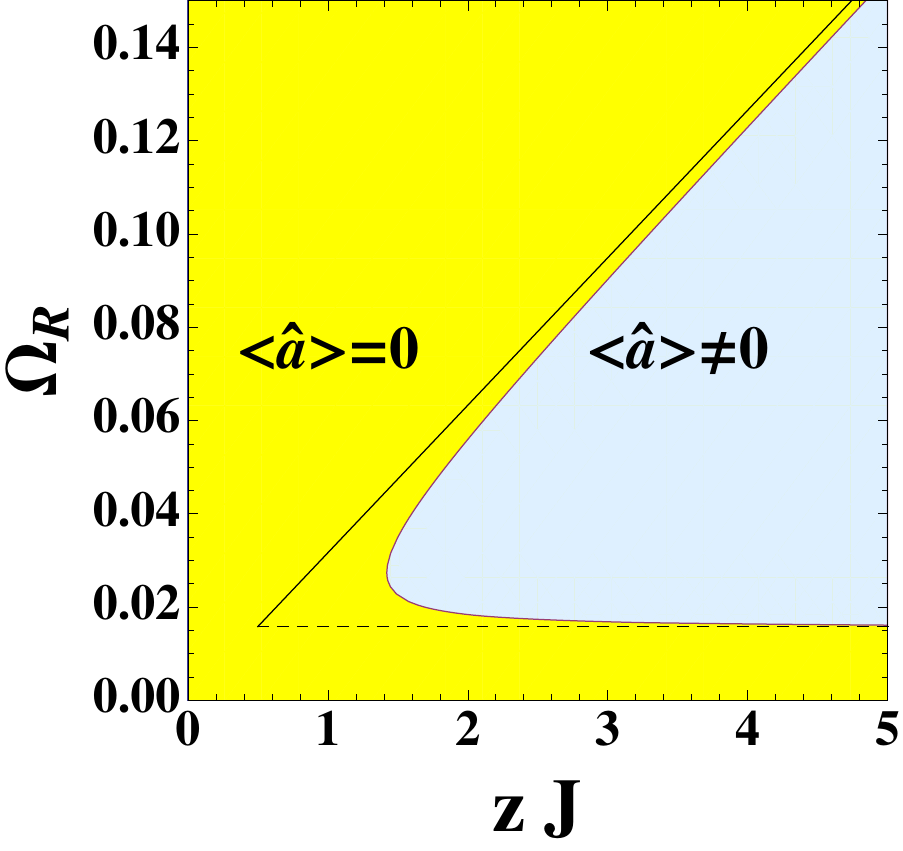}} \\
\subfloat[][]
{\includegraphics[width=.47\columnwidth]{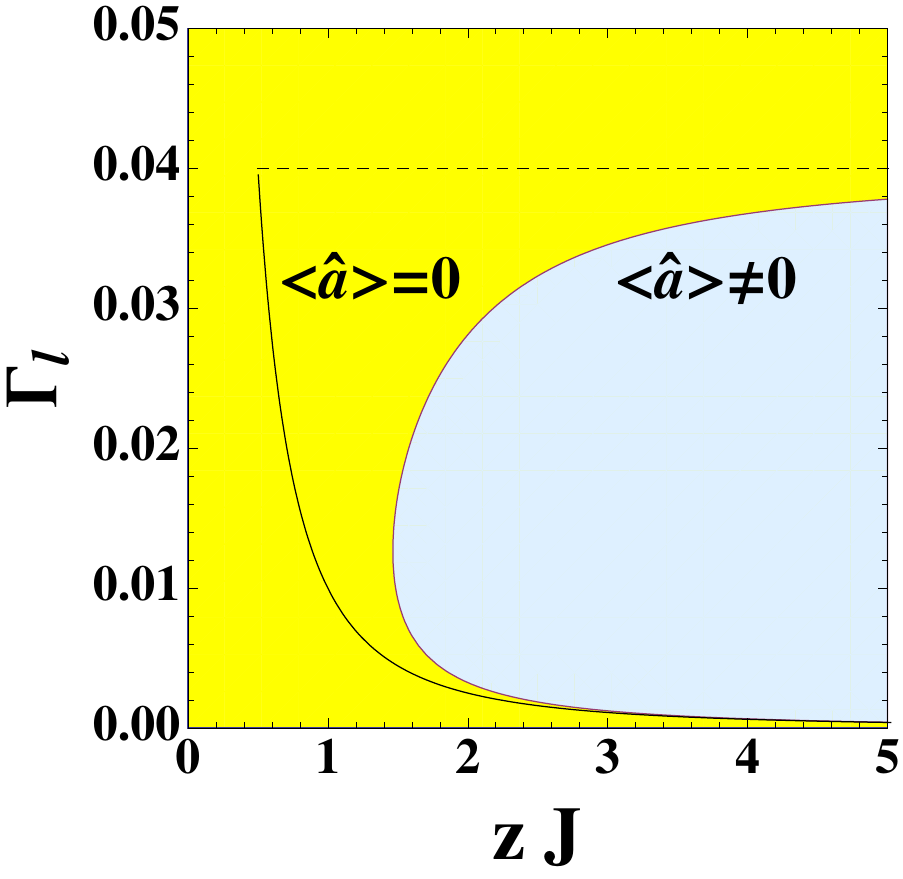}} \quad
\subfloat[][]
{\includegraphics[width=.47\columnwidth]{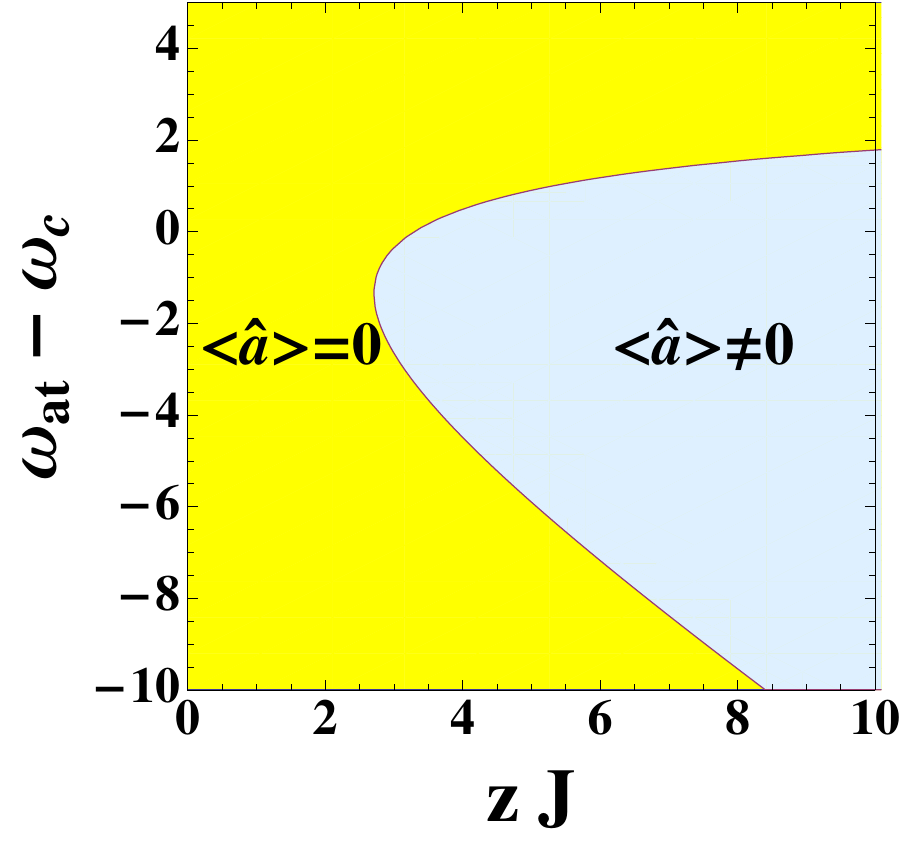}} 
\caption{(Color online) Results of the Gutzwiller mean-field stability analysis in the hard-core regime ($U/J=+\infty$). The yellow area denotes the region where $\braket{\aaa}=0$ is stable, while the light blue area is the region where  the solution $\braket{\aaa}=0$ is unstable and the symmetry is broken.
When not varying them, we fixed the parameters as $\Gamma_p=1$, $\Omega_R=10^{-1}$, $\Gamma_l=\gamma=10^{-3}$, $\omega_{at}=\omega_c$.
The solid lines are the predictions for the critical hopping rate given by Eq.~\eqref{eqstab} which well approximates the phase boundary in the $\Gamma_{em}^0/\Gamma_l\gg1$ limit (see Eq.~\eqref{limstab2}).
The dashed horizontal lines denotes the $\Gamma_{em}^0/\Gamma_l=1$ threshold. The condition $\Gamma_{em}^0/\Gamma_l>1$ is necessary in order to have a significant population in the symmetry broken phase.}
\label{fig:MF_stab}
\end{figure}

In the hard-core regime, it is also possible to exploit the single-cavity exact solution to infer the structure of the phase diagram.
In the unbroken symmetry phase ($\braket{\aaa}=0$), the MF master equation~\eqref{lindblad_cmf} always admits a stationary solution which corresponds to the single-cavity steady-state. 
For certain values of the parameters such solution becomes unstable and the system approaches a stationary state with $\braket{\hat a}\neq0$.
The stability analysis reveals that, for the typical values of the parameters we are interested in, the single-cavity fixed point gets unstable as the hopping rate is increased (see Fig.~\ref{fig:MF_stab}).
The scaling of the instability point with the system parameters can be estimated by comparing the effective emission rate at the band boundary with the photon leakage rate
\be
\Gamma_{em}^{0} \ \frac{\left(\Gamma_p/2\right)^2}{\left(\omega_{at}-\omega_c+z J_c^{\rm HC}\right)^2+\left(\Gamma_p/2\right)^2 }\simeq\Gamma_l.
\ee
In the regime where the effective pumping rate dominates over the photon losses, that is
\be
\label{limstab2}
\frac{\Gamma_{em}^0}{\Gamma_l}=\frac{4 \Omega_R^2}{\Gamma_p\Gamma_l}\gg1,
\ee
at resonance ($\omega_{at}=\omega_c$) we get
\be
\label{eqstab}
z J_c^{\rm HC}  \simeq \sqrt{  \frac{\Gamma_{p}}{\Gamma_{l}}   } \ \Omega_R.
\ee
The prediction of Eq.~\eqref{eqstab} (solid black lines in Fig.~\ref{fig:MF_stab}) well reproduces the phase boundaries in the regime~\eqref{limstab2}.
The dashed horizontal lines in Fig.~\ref{fig:MF_stab} denote the $\Gamma_{em}^0/\Gamma_l=1$ threshold.
The {\it lasing} condition $\Gamma_{em}^0/\Gamma_l\ge1$ is necessary in order to have a significant population in the symmetry broken phase. 
For the details about the Gutzwiller mean-field stability analysis see App.~\ref{app:stab}. 

\begin{figure}[t!]
\centering
\includegraphics[width=.75\columnwidth]{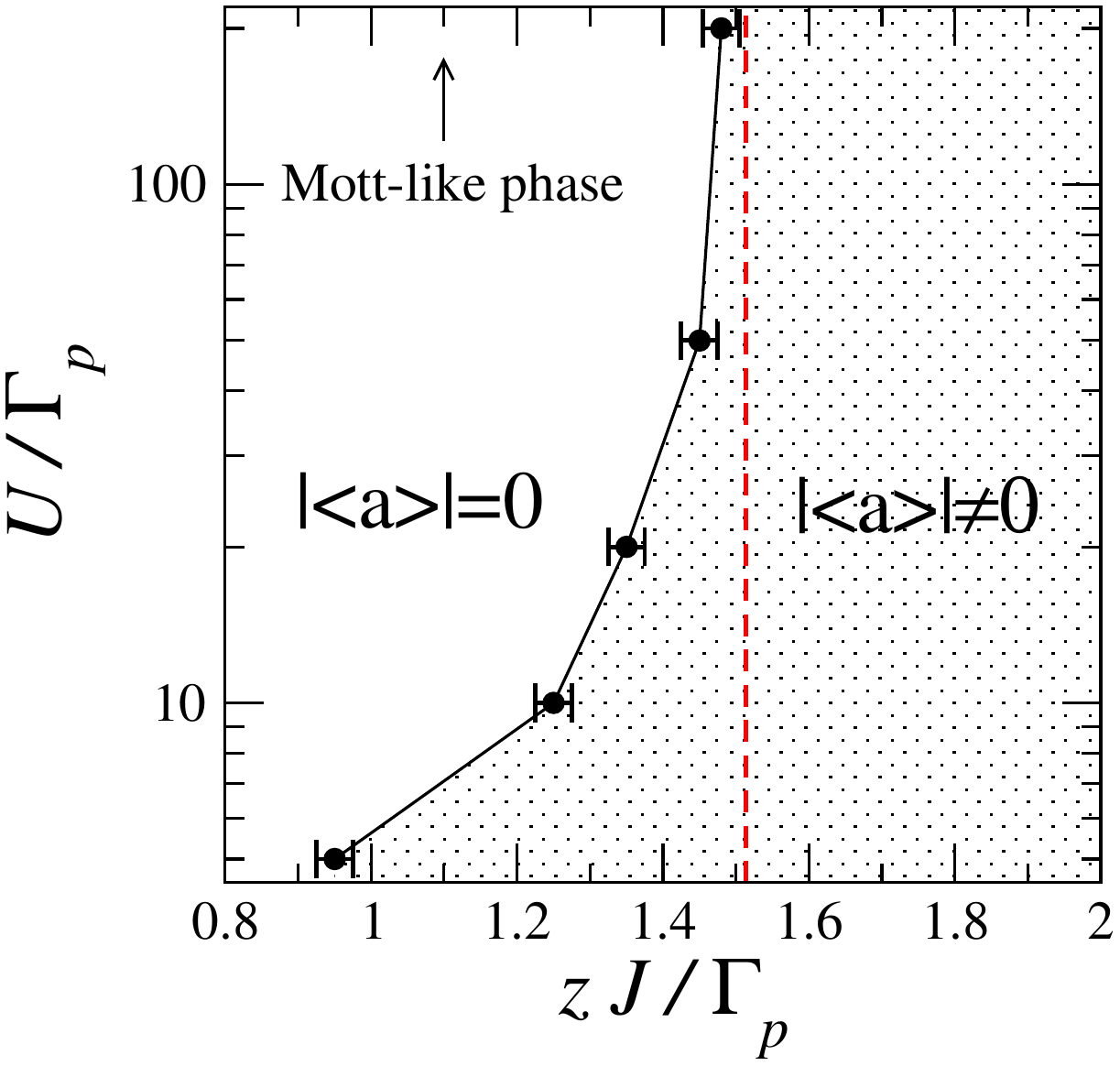} 
\caption{(Color online) Mean-field steady-state phase diagram in the $U/\Gamma_{p}-zJ/\Gamma_{p}$ plane.
The white area corresponds to the region of the parameters for which $|\braket{\aaa}|=0$ while in the dotted region the $U(1)$ symmetry is spontaneously broken and the steady-state exhibits limit-cycle ($|\braket{\aaa}|\neq0$).
Here $\gamma/\Gamma_{p}=\Gamma_{l}/\Gamma_{p}=10^{-2}$, $\Omega_R/\Gamma_{p}=10^{-1}$ and $\omega_{at}=\omega_{c}$.
The dashed vertical line denotes critical hopping rate predicted by the Gutzwiller stability analysis $z J_{c}^{\rm HC}/\Gamma_p= 1.51$.
}
\label{fig:mf_pd}
\end{figure}

It is interesting to extend our study considering finite values of the nonlinearity.
In Fig.~\ref{fig:mf_pd} we show the mean-field steady-state phase diagram in the $U/\Gamma_p-zJ/\Gamma_p$ plane.
For each value of $U$ we explored, we found a critical value of the tunneling rate $J_{c}(U)$ such that for $J>J_{c}(U)$ the $U(1)$ symmetry is spontaneously broken ($|\braket{\aaa}|\neq0$) and the system exhibits limit cycles.
The value of the critical hopping rate increases as $U$ is increased, approaching the value predicted in the hard-core limit $z J_{c}^{\rm HC}/\Gamma_p= 1.51$  (predicted by the Gutzwiller stability analysis).

\subsection{Beyond the Gutzwiller approximation}
\label{sec:results_ex}

In this section, we go beyond the Gutzwiller mean-field approximation employed in Sec.~\ref{ssec:gutz} and present finite-size simulations using different numerical methods. 
The goal is to show what is the fate of the phases predicted by the Gutzwiller mean-field theory.
Since the Hilbert space dimension increases exponentially with the system size, the exact diagonalization of the full Liouvillian becomes soon impracticable.  Numerically, the present problem is also particularly challenging since there are several time scales which differ by orders of magnitude (in particular, the two-level and cavity dissipation rates compared to the incoherent driving rate).
In order to compute the steady-state $\rho^{SS}$ of Eq.~\eqref{lindblad} we use two methods: an algorithm based on the matrix product operators (MPO) formalism~\cite{vestraete2004,zwolak2004} and the corner-space renormalization method~\cite{finazzi2015}.
The MPO algorithm enables us to simulate relatively large chains of cavities (up to $20$ sites) but is unable to explore the region where
long-range spatial correlations develop (convergence with respect to the bond dimension was not achieved in that region). The MPO approach has been therefore used extensively to investigate the nature of the strongly localized Mott-like phase in large arrays.  The corner-space renormalization approach allows us to compute the global density matrix with arbitrary accuracy across the critical region (and therefore can be used to access information as the entropy of the steady state), but it is more limited by the system size (we report here results across the critical region for chains with size up to $8$ sites).

\begin{figure}[b!]
\centering
\includegraphics[width=1.02\columnwidth]{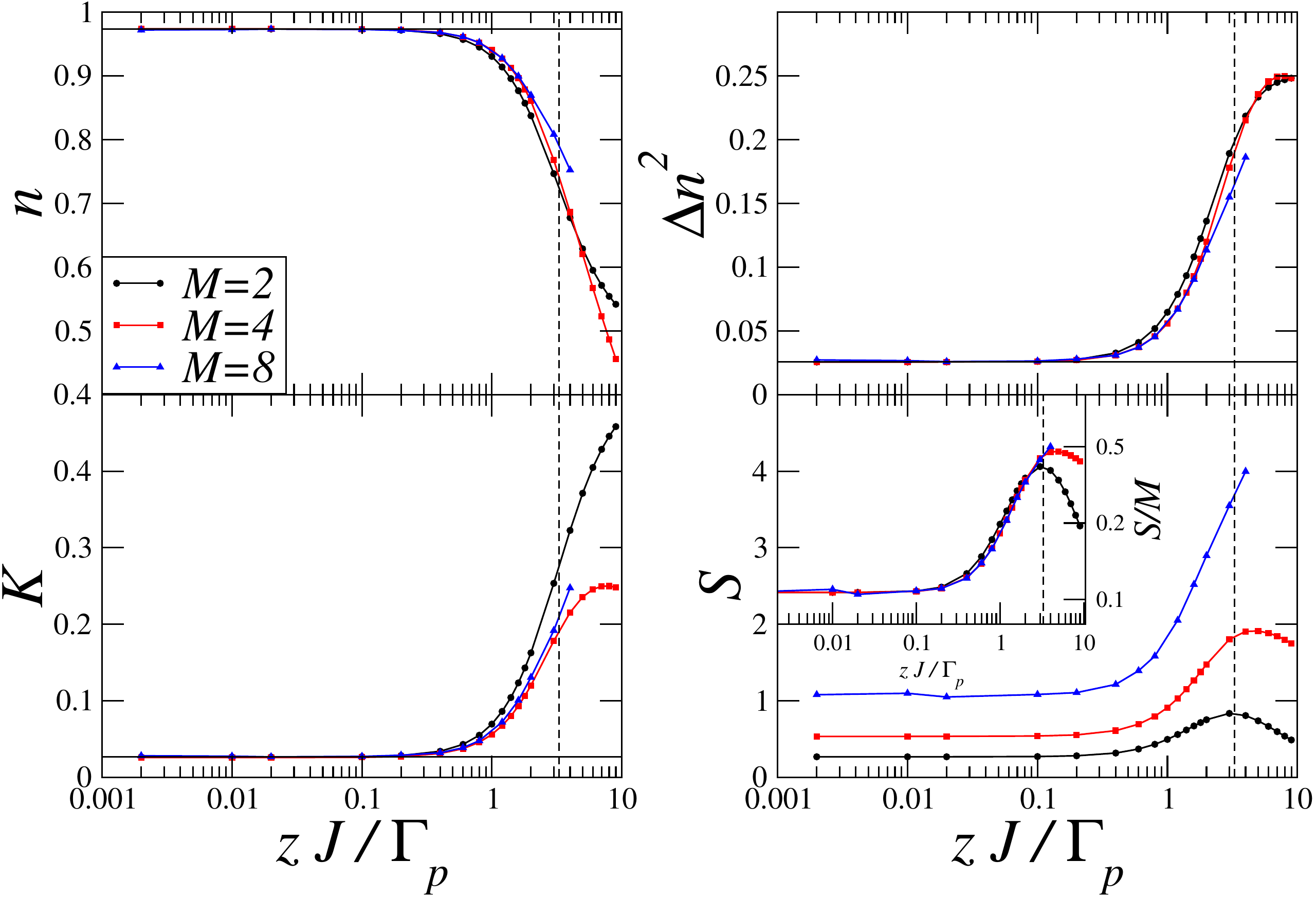}
\caption{(Color online) Top panels: the average photon density $n$ in the steady-state (left panel) and its variance $\Delta n^2$ (right panel) as a function of $J/\Gamma_{p}$.
Bottom panels: steady-state value of the compressibility $\mathcal{K}$ (left panel) and of the entropy (right panel) as a function of $J/\Gamma_{p}$ in the hard-core limit ($U/J=+\infty$). 
The various curves are for different sizes, as indicated in the legend.
The solid horizontal lines are the single-cavity values ($J=0$) of the quantity under consideration.  
The dashed vertical lines denote the critical hopping rates predicted by the Gutzwiller mean-field theory.
The parameters are set as $\omega_{at}=\omega_c-zJ$, $\Gamma_l/\Gamma_p=\gamma/\Gamma_p=10^{-3}$, $\Omega_R/\Gamma_{p}=10^{-1}$.
For the largest size considered ($M=8$), the convergence of the considered quantities has been achieved with $3000$ states in the corner space (the full Hilbert space has a dimension equal to $4^8 = 65536$).}
  \label{fig:1D_detuned_corner_sum}
\end{figure}

In Fig.~\ref{fig:1D_detuned_corner_sum} we show the average photon density $n=\sum_{i=1}^M\braket{\hat a^\dagger_i \hat a_i}/M$, its fluctuations $\Delta n^2=\sum_{i=1}^M\Delta n_i^2/M$, and the compressibility $\mathcal{K}=\sum_{i=1}^M\Delta n_i^2/(n_i M)$ for different system sizes ($M$) as a function of the hopping rate $zJ/\Gamma_{p}$ (here $z=2$) in the hard-core limit. 
In these numerical calculations, we have chosen a slightly shifted atomic frequency $\omega_{at}=\omega_c-zJ$. Under this condition, gain is strongest at the bottom of the photon band, so that condensation into the $\vec{k}=0$ mode is explicitly favored from the outset. This ensures that the spatially homogeneous condensation process that we have assumed in the mean-field calculation is not disturbed by mode competition phenomena between pairs of modes with opposite wavevector which display the same gain and could give rise to condensate fragmentation effects~\cite{frag}. This fragmentation mechanism was likely the reason of the reduced coherence numerically found in~\cite{rivas2014}.

We observe that the data collapse on the single-cavity predictions ($J=0$) for $J\lesssim J_{c}^{\rm HC}$.
This result is in agreement with the findings of the mean-field analysis which predicts a region where local interactions dominate over cooperative effects.
Such Mott-like phase (highlighted by the logarithmic scale) is thus characterized by an almost integer local density and almost vanishing fluctuations and compressibility.
This means that in this parameter range, the correlations among different cavities are very small and cooperative effects are suppressed.
As the photon hopping rate is increased ($J\gtrsim J_{c}^{\rm HC}$), the system enters a novel regime with finite (significant) local density fluctuations and consequently a large value of the compressibility.
Our numerical results in one-dimensional systems are consistent with the expectation that in 1D a {\it true} phase transition to a {\it lasing} regime, as predicted by the Gutzwiller theory, is replaced by a crossover from a Mott-like phase to a mixture of extended Tonks-Girardeau states with different number of photons~\cite{girardeau1960,carusotto2009}.
We remark that the hard-core regime forbids the multiple occupation of a lattice site but allows for the multiple (eventually macroscopic) occupation of a given $\vec{k}$-mode of the lattice.

\begin{figure}[t!]
\centering
\includegraphics[width=0.75\columnwidth]{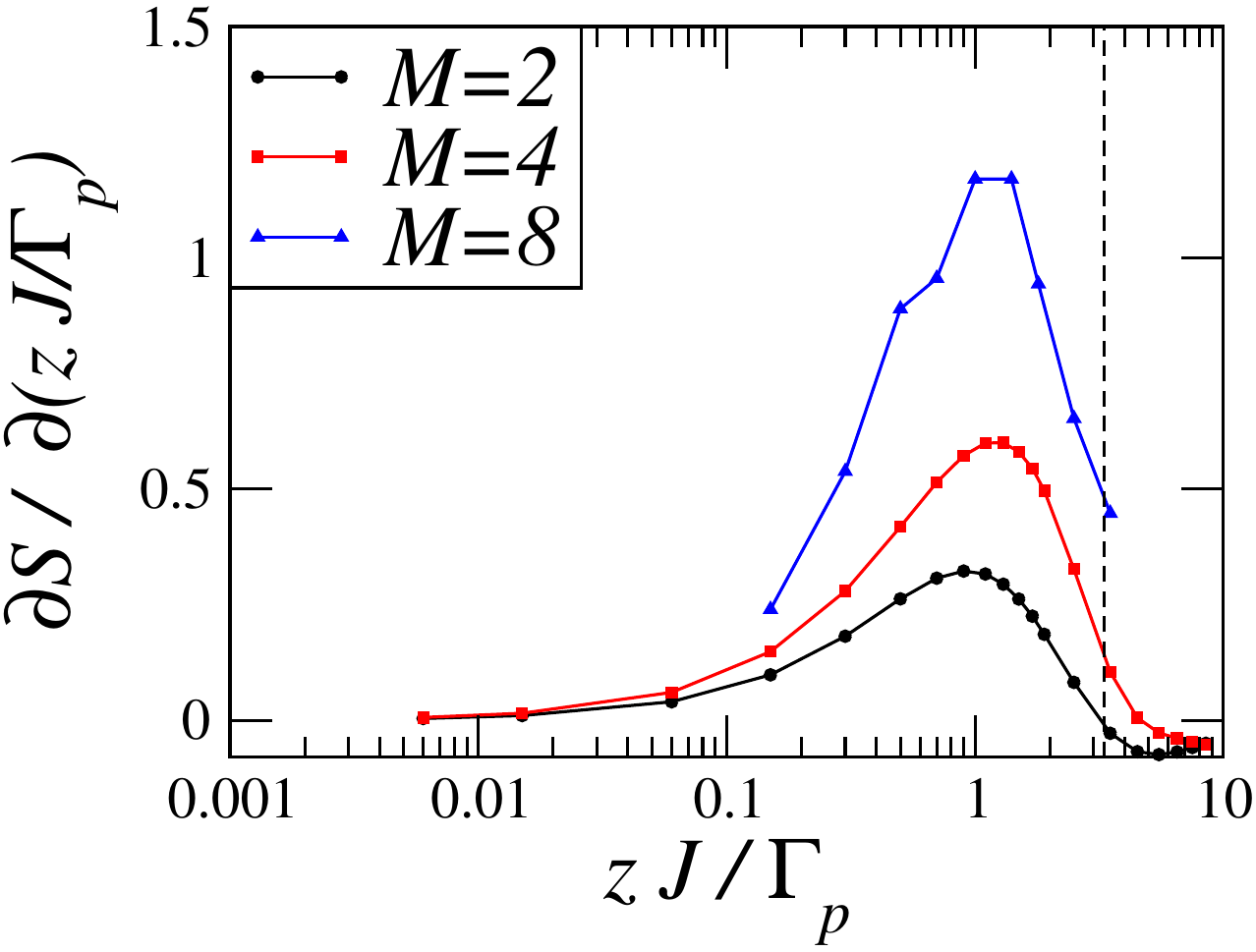}
\caption{(Color online) Derivative of the entropy with respect to the hopping rate $\partial S/\partial(zJ/\Gamma_p)$ for different system sizes as indicated in the legend.
The dashed vertical lines denote the critical hopping rates predicted by the Gutzwiller mean-field theory.
The other parameters are set as in Fig.~\ref{fig:1D_detuned_corner_sum}.}
  \label{fig:1D_detuned_entrderiv}
\end{figure}

The transition between these two phases is also signaled by a sharp variation of the entropy of the steady-state density matrix ($S={\rm Tr}[\rho^{SS}\ln(\rho^{SS})]$, lower right panel of Fig.~\ref{fig:1D_detuned_corner_sum}).
As predicted at mean-field level, in the Mott-like phases the system is an almost pure state while the symmetry broken phase has a strongly mixed character.
This appears as a common feature of phase transitions in driven-dissipative lattices~\cite{jin2016,rota2016}. 
In the inset, we show $S/M$, the entropy rescaled by the number of cavities.
In the Mott-like phase, where the correlations among the cavities are very weak, 
the data referring to different system sizes collapse, indicating the extensive nature of the entropy.
Indeed, when the steady-state density matrix is factorizable one gets $S= M \ S_{\rm sc}$, where $S_{\rm sc}$ is the single cavity entropy.

\begin{figure}[t!]
\centering
\includegraphics[width=0.99\columnwidth]{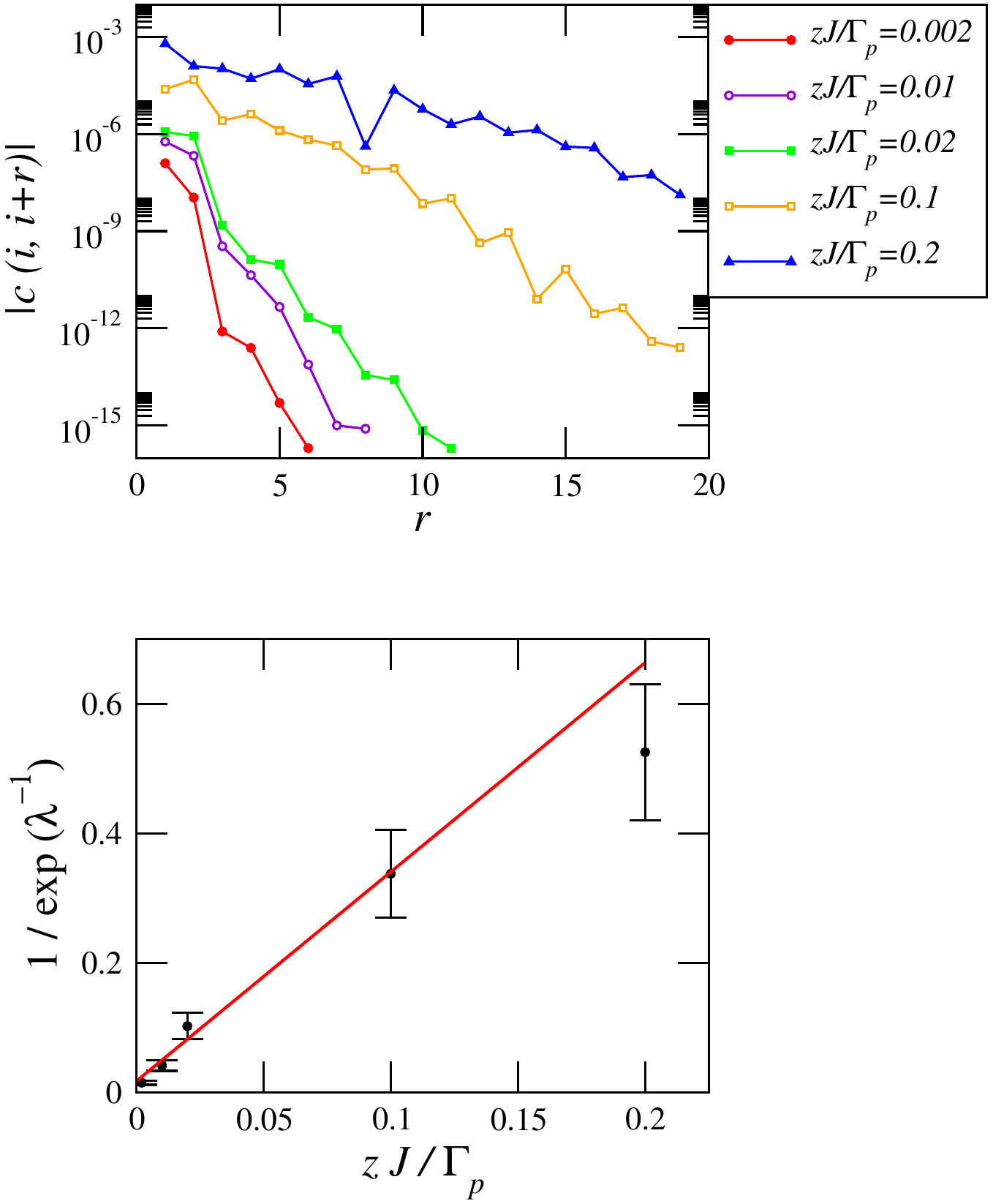}
\caption{(Color online) Top panel: Spatial decay of the correlation function $c(i,i+r)$ (as defined in Eq.~\eqref{corr_fun}) with the distance $r$ for $M=20$, $\omega_{at}=\omega_c$ and different values of $zJ/\Gamma_{p}$, as indicated in the legend. 
Correlators have been chosen in a symmetric way with respect to the center of the chain.
The other parameters are set as in Fig.\ref{fig:1D_detuned_corner_sum}.
Lower panel: The correlation length $\lambda$ obtained fitting $c(i,i+r)$ with an exponentially decaying function. The solid line is the scaling predicted by Eq.~\eqref{scaling_onebody}.
For all the values of the parameters considered, convergence has been achieved with a bond link dimension $\chi=50$ for the MPO algorithm.}
  \label{fig:corr_large_M20}
\end{figure}

In Fig.~\ref{fig:1D_detuned_entrderiv}, we show $\partial S/\partial(zJ/\Gamma_p)$, the derivative of the entropy with respect to the hopping rate: it displays a peak that, for the considered range of sizes,  becomes more pronounced and closer to the Gutzwiller mean-field critical coupling ($zJ_c^{\rm HC}/\Gamma_p=3.3$) as the number of sites is increased. However, it is expected that for one-dimensional systems in the thermodynamic limit no divergence of the entropy derivative occurs since the phase transition predicted by the mean-field theory should be replaced by a crossover between the two phases~\cite{wouters2006,szy2006}. 

To better understand the build up of the quantum correlations among the different cavities 
and to further characterize the Mott-like phase we also computed the one-body correlation function in the steady-state
\begin{equation}
\label{corr_fun}
c(i,j) = \braket{\hat a^\dagger_i \hat a_j}.
\end{equation}
To do so we exploited an MPO algorithm which allows to dynamically reach the steady-state for large systems in the Mott-like regime.
In Fig.~\ref{fig:corr_large_M20} we show the behavior of $c(i,i+r)$ for $M=20$ as $zJ/\Gamma_p$ is varied. 
Since $c(i,j)$ has an oscillatory behaviour due to the very large nonlinearity, we considered its absolute value.
The one-body correlation function decays exponentially with the distance 
\be
\label{corr_fun_02}
|c(i,i+r)| \propto \rme^{-r/\lambda},
\ee
where the correlation length $\lambda$ increases as the photon hopping increases.
In particular, it is possible to show that $\lambda$ scales as 
\be
\label{scaling_onebody}
\lambda \propto \frac{1}{|\ln(J/\Gamma_p)|},
\ee
in the $J/\Gamma_p\ll1$ limit.
The details about the derivation of Eq.~\eqref{scaling_onebody} can be found in App.~\ref{app:onebody}.

\section{Conclusions}
\label{sec:conclu}
In this work, we have investigated the steady-state phases of a photonic lattice in presence of incoherent driving, dissipation and strong photon-photon interactions. We have explored a general model where each lattice site (a nonlinear Kerr resonator) is coupled to a two-level emitter, which is pumped incoherently.
Via a Gutzwiller decoupling theory, we have determined the non-equilibrium phase diagram.
We have found that the interplay between on-site interactions, photon hopping and driven-dissipative processes, lead to a second-order nonequilibrium phase transition related to the spontaneous breaking of the $U(1)$ symmetry possessed by the model.
Furthermore, we have shown that such incoherent driving scheme allows to stabilize Mott-like phases of light characterized by an almost integer local density and almost vanishing compressibility.
The picture predicted by the Gutzwiller mean-field has been validated by numerical finite-size simulations using matrix product operators and the corner-space renormalization method. By driving the system across the critical point,  we have characterized the transition both in terms of one-body correlation functions (which display long-range order) and of the entropy of the system (which increases in the symmetry broken phase).
The phase transition from an (almost) incompressible Mott-like photon fluid to a coherent delocalized phase is driven by commensurability effects.   
The control parameter of this transition can be thus deduced by comparing the effective emission rate at the band boundary with the photon leakage rate.
The mixed character of the steady-state density matrix reflects the intrinsic nonequilibrium nature of the coherent phase. 
Remarkably, signatures of the phase-transition are present already in small arrays. 

The strongly correlated photon phases proposed here could be explored by using photonic quantum simulators based on circuit QED lattices~\cite{houck2012}. In particular, in these systems thanks to superconducting quantum resonators and Josephson junctions it is possible to engineer large Kerr photon-photon interactions and to tailor the interaction with two-level emitters~\cite{wallraff2004,fink2008}, paving the way to the study of strongly correlated quantum fluids of light and non-equilibrium phase transitions. 
The incoherent pump scheme we exploit can be implemented by coherently driving the emitter into a third metastable level from which it fast decays into the excited state of the active transition, thus resulting into an effective incoherent pump~\cite{ma2017}.
The role of dimensionality, disorder and criticality in these nonequilibrium quantum phases are intriguing topics that need to be explored experimentally and theoretically in the future.

\section{Acknoledgments}
Discussions with Leonardo Mazza are warmly acknowledged.
AB, FS and CC acknowledge support from ERC (via Consolidator Grant CORPHO No. 616233). 
AB, CC, DR, FS, IC and RF acknowledge the Kavli Institute for Theoretical Physics, University of California, Santa Barbara (USA) for the hospitality and support during the early stage of this work.
We acknowledge the CINECA award under the ISCRA initiative, for the availability of high performance computing resources and support.
JL and IC are supported by the EU-FET Proactive grant AQuS, Project No. 640800, and by the Autonomous Province of Trento, partially through the project ``On silicon chip quantum optics for quantum computing and secure communications'' (``SiQuro'').
RF acknowledges Oxford Martin School Technologies  and the Singapore Ministry of Education and National Research Foundation (QSYNC) for financial support.


\appendix

\section{Single-cavity solution (hard-core limit)}
\label{app:hcsol}
When the nonlinearity is large enough to forbid the double occupation of a lattice site, one can map the bosonic degree of freedom into an effective spin
\ba
\aaa &\to& \hat{\Sigma}^- \cr
&&\cr
\ada &\to& \hat{\Sigma}^+.
\ea
Doing so, the single-cavity Hamiltonian becomes
\be
\label{effham_spin}
\hat{H} =\frac{\omega_c}{2} (\sssz+1) + \frac{\omega_a}{2} (\ssz+1) + \frac{\Omega_R}{2} (\sssx\ssx+\sssy\ssy).
\ee
In the Heisenberg representation the evolution of a given operator $\hat{\theta}$ is ruled by
\be
\label{mf_heis}
\dot\theta = \rmi [\hat{H},\hat\theta] + \mathcal{\tilde{L}}[\rho],
\ee 
where
\be
\mathcal{\tilde{L}}[\rho] = \sum_{i} \left( \frac{\Gamma_{l}}{2} \mathcal{\tilde{D}}[\aaa_i;\rho] + \frac{\gamma}{2} \mathcal{\tilde{D}}[\hat\sigma_i^{-};\rho] + \frac{\Gamma_{p}}{2} \mathcal{\tilde{D}}[\hat\sigma_i^{+}; \rho]\right), 
\ee
with $\mathcal{\tilde{D}}[\hat{O};\rho] = [ 2\hat{O}^\dagger \rho\hat{O}  -\{   \hat{O}^\dagger  \hat{O},\rho  \} ]$.
Writing the equation of motion for $\braket{\sssa}$, $\braket{\ssa}$, $\braket{\sssa \hat{\sigma}^\beta}$ ($15$ equations) and solving for the stationary state one always find a single (stable) solution given by

\ba
\label{mf_heis_sol}
\braket{\sssx}&=&\braket{\sssy}=\braket{\ssx}=\braket{\ssy}=0, \cr
&&\cr
&&\cr
\braket{\sssz}&=&
\frac{-\Gamma_{l} (\Gamma_{p}+\gamma) \Lambda+4\Omega_R^2 (\Gamma_{p}^2-(\gamma+\Gamma_{l})^2) }{\Gamma_{l} (\Gamma_{p}+\gamma) \Lambda+4\Omega_R^2 \Gamma_{tot}^2},\cr
&&\cr
&&\cr
\braket{\ssz}&=&
\frac{\Gamma_{l}  (\Gamma_{p}-\gamma) \Lambda+ 4\Omega_R^2 (\Gamma_{p}^2-(\gamma+\Gamma_{l})^2)}{\Gamma_{l} (\Gamma_{p}+\gamma) \Lambda+4\Omega_R^2 \Gamma_{tot}^2}, \cr
&&\cr
&&\cr
\braket{\sssx\ssz}&=&\braket{\sssy\ssz}=\braket{\sssz\ssx}=\braket{\sssz\ssy}=0,\cr
&&\cr
&&\cr
\braket{\sssx\ssx}&=&\braket{\sssy\ssy}=
\frac{8 \Gamma_{p} \Gamma_{l}  \Omega_R \Delta\omega}{\Gamma_{l} (\Gamma_{p}+\gamma) \Lambda+4\Omega_R^2 \Gamma_{tot}^2},\cr
&&\cr
&&\cr
\braket{\sssx\ssy}&=&-\braket{\sssy\ssx}=
-\frac{4\Gamma_{p}\Gamma_{l} \Omega_R  \Gamma_{tot}}{\Gamma_{l} (\Gamma_{p}+\gamma) \Lambda+4\Omega_R^2 \Gamma_{tot}^2}, \cr
&&\cr
&&\cr
\braket{\sssz\ssz}&=&
\frac{4\Omega_R^2 (-\Gamma_{p}+\gamma+\Gamma_{l})^2-\Gamma_{l} (\Gamma_{p}-\gamma) \Lambda}{\Gamma_{l} (\Gamma_{p}+\gamma) \Lambda+4\Omega_R^2 \Gamma_{tot}^2},
\ea
where $\Delta\omega=\omega_{at}-\omega_{c}$, $\Gamma_{tot}=\Gamma_{p}+\gamma+\Gamma_{l}$ and $\Lambda=\Gamma_{tot}^2+4 \Delta\omega^2$.

\section{Mean-field stability analysis (hard-core limit)}
\label{app:stab}

At mean-field level the dynamics in the hard-core regime is ruled by the master equation~\eqref{mf_heis} where the single-cavity Hamiltonian~\eqref{effham_spin} is replaced by
\be
\hat{H} \to \hat{H} - \frac{z J}{2} (\sssx\braket{\sssx}+\sssy\braket{\sssy}).
\ee
The single-cavity steady-state solution~\eqref{mf_heis_sol} is always a fixed point of the single-site MF equations with $\braket{\aaa}=0$ ($\braket{\sssx}=\braket{\sssy}=0$) which becomes unstable for certain values of the parameters and allows the emergence of a phase characterized by $\braket{\aaa}\neq0$.
In order to study the stability of this solution we computed the Jacobian matrix of the MF equations~\cite{leboite2013,lee2013,wilson2016} evaluated with respect to the single-cavity solution~\eqref{mf_heis_sol}.
If the real part of one of its eigenvalues becomes positive for certain values of the parameters then the single-cavity fixed point is unstable.
In Fig.~\ref{fig:MF_stab} we show the Heaviside step function of the real part of the most unstable eigenvalue of the Jacobian $\Theta\left[  {\rm Re}\left( \lambda_u \right)\right]$ as a function of the typical values of the system parameters.
The result is a plot which highlights the stable ($\braket{\aaa}=0$) and the unstable ($\braket{\aaa}\neq0$) regions in yellow and light blue respectively.

\section{Spatial correlation functions (1D)}
\label{app:onebody}

In this appendix we provide a scaling law for the spatial decay of the one-body correlations in the $J/\Gamma_{p}\ll1$ regime by using an ansatz for the photonic steady-state density matrix in the hard-core limit.
Even if the derivation might at first appear somewhat {\it heuristic}, it will provide a good approximation of how the correlation length depends on the system parameters (see Sec.~\ref{sec:results_ex}).  
In analogy with the equilibrium physics of hard-core bosons in one dimension, we suppose that the steady-state is fermionized, i.e., that the photonic density matrix $\rho$ is diagonal in the fermionic momentum basis, up to a unitary Jordan-Wigner transformation $\hat{U}$ which anti-symmetrizes the bosonic density matrix:
\begin{equation}
\rho^{F}=\hat{U} \rho^{B} \hat{U}^{-1}, \quad \rho^{F}=\otimes_k\: \rho^{F}_{k}.
\end{equation}
The bosonic and fermionic annihilation operators are related through the unitary relation
\begin{equation}
\aaa_{j}^{F}=\rme^{\rmi\pi(\sum_{l<j} \hat{n}_{l})}\hat{U} \aaa_{j}^{B} \hat{U}^{-1},
\end{equation}
where the local particle number operator $\hat{n}_{l}=\hat{n}^{B/F}_{l}$ is left unchanged by  the anti-symmetrization process.

In the simple case of free bosons, the steady state momentum distribution can be exactly calculated analytically:
\begin{eqnarray}
\label{eq:free-bosons}
n^{fb}_k &=&\left(\frac{\Gamma_{l}}{\Gamma_{em}(\omega_c-\epsilon_k)}-1\right)^{-1}\nonumber\\
&& \cr
&=&\left(\frac{\Gamma_{l}}{\Gamma_{em}^0}\left[\left(\tilde{\delta}+2\epsilon_{k}/\Gamma_{p}\right)^2+1\right]-1\right)^{-1},
\end{eqnarray}
where $\tilde{\delta}=-2 \Delta\omega/\Gamma_{p}$ and $\epsilon_k=-2J\cos(k)$. So that, a natural ansatz for the Fermi non-equilibrium distribution would be
\begin{equation}
\label{eq:free-fermions}
n^{ff}_k=\left(\frac{\Gamma_{l}}{\Gamma_{em}^0}\left[\left(\tilde{\delta}+2\epsilon_{k}/\Gamma_{p}\right)^2+1\right]+1\right)^{-1}.
\end{equation}
The analytical function 
\be
g(z)=\left(\frac{\Gamma_{l}}{\Gamma_{em}^0}\left[\left(\tilde{\delta}+z\right)^2+1\right]+1\right)^{-1}
\ee 
of the complex variable $z$ can be expanded as
\begin{equation}
\label{eq:serie}
g(z)=\sum_{n}\alpha_{n}(z/r_{c})^n,
\end{equation}
where $r_{c}$ is the convergence radius of this power serie, and $\alpha_{n}$ is a sub-exponential function of the variable $n$ which is dominated by any geometric function. The convergence radius is given by the modulus of the complex pole of $g(z)$ which is closest to the origin $z=0$, and thus has the following expression 
$r_{c}=\sqrt{1+\tilde{\delta}^2+\Gamma_{em}^0/\Gamma_{l}}$.

From the ansatz Eq.~\eqref{eq:free-fermions}, it is possible to calculate the long-range properties of the fermionic one-body correlation function
\begin{equation}
\label{eq:correlation-fermion}
c^{F}(j)=\braket{\hat{a}^{F}_{j}\hat{a}_{0}^{F\dagger}}=\int^{\pi}_{-\pi}\frac{dk}{2\pi}e^{ik j} \ n^{ff}_k.
\end{equation}
Exploiting the expansion Eq.~\eqref{eq:serie} we obtain
\begin{eqnarray}
\label{eq:correlation-fermion}
c^{F}(j)&=&\sum_{n\geq j} \alpha_{n} \left(\frac{2J}{ r_{c}\Gamma_{p}}\right)^n \int^{\pi}_{-\pi}\frac{dk}{2\pi}e^{ikj}\text{cos}^{n}(k),
\end{eqnarray}
as the integral in the right side is non-zero only for $n\geq j$. In the limit, $2J/ (r_{c}\Gamma_{p})\ll1$ we can keep only the lowest power of $J$ in the serie expansion, which corresponds to $n=j$
\begin{eqnarray}
\label{eq:correlation-fermion}
c^{F}(j)&\simeq &\alpha_{j} \left(\frac{2J}{ r_{c}\Gamma_{p}}\right)^j \int^{\pi}_{-\pi}\frac{dk}{2\pi}e^{ik j}\cos^j(k)\\
&& \cr
&= &\alpha_{j} \left(\frac{J}{ r_{c}\Gamma_{p}}\right)^j\\
&& \cr
&=&\alpha_{j} \ e^{-j/\lambda_{F}},
\end{eqnarray}
where $\alpha_{j}$ is sub-exponential. The fermionic autocorrelation thus spatially decays exponentially with a correlation length $\lambda_F$ given by
\begin{equation}
\label{eq:length-fermions}
1/\lambda_{F}=\text{ln}\left(\frac{\Gamma_{p}\sqrt{1+\tilde{\delta}^2+\Gamma_{em}^0/\Gamma_{l}}}{J}\right),
\end{equation}
and scales as $1/\text{ln}(\Gamma_{p}/J)$ for small $J/\Gamma_{p}$. 

The true photonic correlation function is related to the fermionic one as follows
\begin{equation}
\label{eq:correlation-bosons}
c(j)=\braket{\hat{a}_{j}\hat{a}_{0}^{\dagger}}=\braket{\hat{a}^{F}_{j}\rme^{\rmi\pi\sum_{0<l<k}\hat{n}_{l}}\hat{a}_{0}^{F\dagger}}.
\end{equation}
For a vanishing $J$, the fermionic distribution is nearly momentum independent, and the particles are thus fully localized with the uniform spatial density $n=1/(\Gamma_{l}/\Gamma_{em}^0+1)$. 
The correlation function can be thus factorized as
$c(j)\simeq \braket{\aaa^{F}_{j}\aaa_{0}^{F\dagger}}\prod_{0<l<k}\braket{e^{i\pi\hat{n}_{l}}}$. 
The local expectation value $\braket{e^{i\pi\hat{n}_{l}}}\simeq (1-\Gamma_{em}^0/\Gamma_{l})/(1+\Gamma_{em}^0/\Gamma_{l})$, is positive for hole-dominated statistics ($\Gamma_{em}^0/\Gamma_{l}<1$) and negative for particle dominated statistics ($\Gamma_{em}^0/\Gamma_{l}>1$). 

The expression for the photonic one-body correlation function in the $J/\Gamma_p\ll1$ limit reads as
\be
\label{eq:correlation-bosons}
c(j) \propto  (-1)^j \ e^{-j/\lambda_{ph}},
\ee
for particle-dominated statistics. 
The corresponding correlation length is given by
\be
\label{eq:length-photon}
1/\lambda_{ph}=\text{ln}\left(\left|\frac{1+\frac{\Gamma_{em}^0}{\Gamma_{l}}}{1-\frac{\Gamma_{em}^0}{\Gamma_{l}}}\right|\frac{\Gamma_{p}\sqrt{1+\tilde{\delta}^2+\frac{\Gamma_{em}^0}{\Gamma_{l}}}}{J}\right).
\ee
It maintains the logarithmic scaling, and is slightly shorter than the fermionic one, due to a scrambling induced by sign changes when crossing intermediary particles.



\begin{thebibliography}{99}
\bibitem{note00}
Let us note that definition of compressibility $\mathcal{K}=\Delta n^2/n$ we use corresponds to a renormalized photon density.
It gets the physical meaning of a compressibility (that is a response in density to a change of chemical potential) only at equilibrium when it is possible to apply a fluctuation-dissipation theorem.
\bibitem{note01}
Let us note that the string-order parameter~\cite{stringorder} used in equilibrium to witness the emergence of a Mott insulator phase in this case is strictly zero since the local density can never be exactly an integer.
\bibitem{note02} 
We recall that since that dynamics is governed by the master equation~\eqref{lindblad} the $U(1)$ symmetry does not lead to any conserved quantity as it would be the case in absence of driving and dissipation. In such a case the $U(1)$ symmetry is related to the conservation of the total number of excitations.
\bibitem{note03}
A richer physics could of course be described within the Gutzwiller approach by letting the local density matrix vary from site to site ($\rho_i\neq\rho_j$).  
The price for this generality would however be a physically more intricate behavior and a much less transparent theory. This could be of interest if one wants to describe, e.g.,  the effect of strong interactions onto mode-competition effects.
\end{thebibliography}
\end{document}